\documentclass[12pt]{article}
\usepackage[dvips]{color}
\usepackage{epsfig}
\usepackage{amsmath}  
\usepackage{graphicx}
\begin{document}
\title{\bf {Interacting new agegraphic viscous dark energy with varying $G$} }
\author{{ A. Sheykhi $^{a,c}$ \thanks{Email: sheykhi@mail.uk.ac.ir}\hspace{1mm}
, M. R. Setare $^{b,c}$\thanks{Email:
rezakord@ipm.ir}}\\
{ $^{a}$ Department of Physics, Shahid Bahonar University,}\\
{P.O. Box 76175, Kerman, Iran}\\{ $^b$ Department of Science,
Payame Noor University, Bijar, Iran}\\ { $^c$ Research Institute
for Astronomy and Astrophysics of Maragha}\\ {(RIAAM), Maragha,
         Iran}\\ }

\date{}

\maketitle
\begin{abstract}
We consider the new agegraphic model of dark energy with a varying
gravitational constant, $G$, in a non-flat universe. We obtain the
equation of state and the deceleration parameters for both
interacting and noninteracting new agegraphic dark energy. We also
present the equation of motion determining the evolution behavior
of the dark energy density with a time variable gravitational
constant. Finally, we generalize our study to the case of viscous
new agegraphic dark energy in the presence of an interaction term
between both dark components.
\end{abstract}
\newpage

\section{Introduction\label{Introd}}
Many cosmological observations, such as SNe Ia \cite{1}, WMAP
\cite{2}, SDSS \cite{3}, Chandra X-ray observatory \cite{4}, etc.,
reveal that our universe is undergoing an accelerating expansion.
To explain this cosmic positive acceleration, mysterious dark
energy has been proposed. There are several dark energy models
which can be distinguished by, for instance, their equation of
state (EoS) $(w=\frac{P_{de}}{\rho_{de}}) $ during the evolution
of the universe. Although the simplest way to explain this
behavior is the consideration of a cosmological constant
\cite{c7}, the known fine-tuning problem \cite{8} led to the dark
energy paradigm. The dynamical nature of dark energy, at least in
an effective level, can originate from various fields, such is a
canonical scalar field (quintessence) \cite{quint}, a phantom
field, that is a scalar field with a negative sign of the kinetic
term \cite{phant}, or the combination of quintessence and phantom
in a unified model named quintom \cite{quintom}.

An approach to the problem of DE arises from the holographic
principle that states that the number of degrees of freedom
related directly to entropy scales with the enclosing area of the
system. It was shown by 'tHooft and Susskind \cite{hologram} that
effective local quantum field theories greatly overcount degrees
of freedom because the entropy scales extensively for an effective
quantum field theory in a box of size $L$ with UV cut-off $
\Lambda$. Attempting to solve this problem, Cohen {\it et al.}
showed \cite{cohen} that in quantum field theory, short distance
cut-off $\Lambda$ is related to long distance cut-off $L$ due to
the limit set by forming a black hole. In other words the total
energy of the system with size $L$ should not exceed the mass of
the same size black hole i.e. $L^3 \rho_{\Lambda}\leq LM_p^2$
where $\rho_{\Lambda}$ is the quantum zero-point energy density
caused by UV cutoff $\Lambda$ and $M_P$ denotes Planck mass (
$M_p^2=1/{8\pi G})$. The largest $L$ is required to saturate this
inequality. Then its holographic energy density is given by
$\rho_{\Lambda}= 3c^2M_p^2/ L^2$ in which $c$ is free
dimensionless parameter and coefficient $3$ is for convenience.
More recently a new dark energy model, dubbed agegraphic dark
energy has been proposed \cite{cai1} (see also \cite{zin}), which
takes into account the Heisenberg uncertainty relation of quantum
mechanics together with the gravitational effect in general
relativity. Following the line of quantum fluctuations of
spacetime, Karolyhazy \cite{kar} proposed that the distance $t$ in
Minkowski spacetime cannot be known to a better accuracy than
$\delta t=\beta t_{p}^{2/3}t^{1/3}$, where $\lambda$ is a
dimensionless constant of order unity. Based on Karolyhazy
relation, Maziashvili proposed that the energy density of metric
fluctuations of Minkowski spacetime is given by \cite{maz}
\begin{equation}\label{2}
\rho_{\Lambda}\sim\frac{1}{t_{p}^{2}t^{2}}\sim\frac{M_{p}^{2}}{t^{2}},
 \end{equation}
 where $t_p$ is the reduced Planck time, and $M_{p}$ is the Planck
 mass. The agegraphic models of dark
energy  have been examined  and constrained by various
astronomical observations \cite{age,shey1,shey2,setare,age2}.

Since we know neither the nature of dark energy nor the nature of
dark matter, a microphysical interaction model is not available
either. However, pressureless dark matter in interaction with
holographic dark energy is more than just another model to
describe an accelerated expansion of the universe. Understanding
dark energy is one of the biggest challenges to the particle
physics of this century. Studying the interaction between the dark
energy and ordinary matter will open a possibility of detecting
the dark energy. It should be pointed out that evidence was
recently provided by the Abell Cluster A586 in support of the
interaction between dark energy and dark matter
\cite{Bertolami:2007zm}. However, despite the fact that numerous
works have been performed till now, there are no strong
observational bounds on the strength of this interaction
\cite{Feng:2007wn}. This weakness to set stringent (observational
or theoretical) constraints on the strength of the coupling
between dark energy and dark matter stems from our unawareness of
the nature and origin of dark components of the Universe. It is
therefore more than obvious that further work is needed to this
direction.

Previously, it has been shown that the interacting new agegraphic
model of dark energy can cross the phantom divide \cite{setare}.
The phantom energy, if it exists, can cause some peculiar
phenomena e.g. violates the strong energy condition, $\rho+3p\geq
0$. This leads us to consider
phantom energy as an imperfect fluid, implying that the phantom
fluid could contain non-zero bulk and shear viscosities
\cite{shee}. The bulk viscosities are negligible for
non-relativistic and ultra-relativistic fluids but are important
for the intermediate cases. In viscous cosmology, shear
viscosities arise in relation to space anisotropy while the bulk
viscosity accounts for the space isotropy \cite{Bre1}. Generally,
shear viscosities are ignored (as the CMBR does not indicate
significant anisotropies) and only bulk viscosities are taken into
account for the fluids in the cosmological context. Moreover, bulk
viscosity related to a grand unified theory phase transition may
lead to an explanation of the accelerated cosmic expansion
\cite{lanc}.\\
Although the holographic dark energy model with varying
gravitational constant has been studied in \cite{jamil,manos},
however, until now, in all the investigated agegraphic dark energy
models a constant Newton's "constant" G has been considered.  The
role of $G$-variation will be expressed through the pure number
$G'/G\equiv\Delta_G $, which will be extracted from observations.
In particular, observations of Hulse-Taylor binary pulsar
B$1913+16$ lead to the estimation
$\dot{G}/G\sim2\pm4\times10^{-12}{yr}^{-1}$ \cite{kogan,Damour},
while helio-seismological data provide the bound
$-1.6\times10^{-12}{yr}^{-1}<\dot{G}/G<0$ \cite{guenther}.
Similarly,  Type Ia supernova observations \cite{1}  give the best
upper bound of the variation of $G$ as $-10^{-11} yr^{-1} \leq
\frac{\dot G}{G}<0$ at redshifts $z \simeq 0.5$ \cite{Gaztanaga},
while astereoseismological data from the pulsating white dwarf
star G117-B15A lead to $\left|\frac{\dot G}{G}\right| \leq 4.10
\times 10^{-11} yr^{-1}$ \cite{Biesiada}. See also \cite{ray1} for
various bounds on $\dot{G}/G$ from observational data, noting that
all these measurements are valid at relatively low redshifts, i.e
$z<3.5$. Since the limits in $G$-variation are given for
$\dot{G}/G$ in units $yr^{-1}$, and since $\dot{G}/G=H G'/G$, we
can estimate $\Delta_G $ substituting the value of $H$ in
$yr^{-1}$ \cite{jamil}.

Besides, there have been many proposals in the literature
attempting to theoretically justified a varying gravitational
constant. For example in  Brans-Dicke theory the gravitational
constant is replaced by a scalar field coupling to gravity through
a new parameter, and it has been generalized to various forms of
scalar-tensor theories \cite{Ber}, leading to a considerably
broader range of variable-G theories. In addition, justification
of a varying Newton's constant has been established with the use
of conformal invariance and its induced local transformations
\cite{Bek}. Finally, a varying G can arise perturbatively through
a semiclassical treatment of Hilbert-Einstein action \cite{Shap},
non-perturbatively through quantum gravitational approaches within
the ``Hilbert-Einstein truncation'' \cite{Reu}, or through
gravitational holography \cite{Horv,Gub}.

In the light of all mentioned above, it becomes obvious that the
investigation on the interacting new agegraphic dark energy models
with varying gravitational constant is well motivated. In this
paper, we would like to generalize, following \cite{set1}, the new
agegraphic viscous dark energy models to the universe with spacial
curvature in the presence of interaction between the dark matter
and dark energy with a varying gravitational constant, $G$.
\section{New ADE with varying gravitational constant \label{NADE}}

Soon after the original ADE model was introduced \cite{cai1}, an
alternative model dubbed `` new agegraphic dark energy" was proposed
by  Wei and Cai \cite{Wei2}, while the time scale is chosen to be
the conformal time $\eta$ instead of the age of the universe, which
is defined by $dt= ad\eta$, where $t$ is the cosmic time. It is
worth noting that the Karolyhazy relation $\delta{t}= \beta
t_{p}^{2/3}t^{1/3}$ was derived for Minkowski spacetime $ds^2 =
dt^2-d\mathrm{x^2}$ \cite{kar,maz}. In case of the FRW universe, we
have $ds^2 = dt^2-a^2d\mathrm{x^2} = a^2(d\eta^2-d\mathrm{x^2})$.
Thus, it might be more reasonable to choose the time scale to be the
conformal time $\eta$ since it is the causal time in the Penrose
diagram of the FRW universe. The new ADE contains some new features
different from the original ADE and overcome some unsatisfactory
points. For instance, the original ADE suffers from the difficulty
to describe the matter-dominated epoch while the new agegraphic dark
energy resolved this issue \cite{Wei2}. The energy density of the
new ADE can be written
\begin{equation}\label{rhon}
\rho_{D}= \frac{3n^2 }{8\pi G\eta^2},
\end{equation}
where the conformal time is given by
\begin{equation}
\eta=\int{\frac{dt}{a}}=\int_0^a{\frac{da}{Ha^2}}.
\end{equation}
If we write $\eta$ to be a definite integral, there will be an
integral constant in addition. Thus, we have $\dot{\eta}=1/a$. Let
us again consider a FRW universe with spatial curvature containing
the new agegraphic dark energy and pressureless matter. The
Friedmann equation can be written
\begin{eqnarray}\label{Friednew}
H^2+\frac{k}{a^2}=\frac{8\pi G}{3} \left( \rho_m+\rho_D \right),
\end{eqnarray}
where $k$ is the curvature parameter with $k = -1, 0, 1$
corresponding to open, flat, and closed universes, respectively. A
closed universe with a small positive curvature
($\Omega_k\simeq0.01$) is compatible with observations \cite{spe}.
If we introduce, as usual, the fractional energy densities such as
\begin{eqnarray}\label{Omega}
\Omega_m=\frac{8\pi G\rho_m}{3H^2}, \hspace{0.5cm}
\Omega_D=\frac{8\pi G\rho_D}{3H^2},\hspace{0.5cm}
\Omega_k=\frac{k}{H^2 a^2},
\end{eqnarray}
then the Friedmann equation can be written
\begin{eqnarray}\label{Fried2}
\Omega_m+\Omega_D=1+\Omega_k.
\end{eqnarray}
The fractional energy density of the new agegraphic dark energy
can also be written
\begin{eqnarray}\label{Omegaqnew}
\Omega_D=\frac{n^2}{H^2\eta^2}.
\end{eqnarray}
We consider the FRW universe filled with dark energy and dust
(dark matter) which evolves according to their conservation laws
\begin{eqnarray}
&&\dot{\rho}_D+3H\rho_D(1+w_D)=0,\label{consq}\\
&&\dot{\rho}_m+3H\rho_m=0, \label{consm}
\end{eqnarray}
where $w_D=p_D/\rho_D$ is the equation of state parameter of new
ADE. Taking the derivative of Eq. (\ref{rhon}) with respect to the
cosmic time and using Eq. (\ref{Omegaqnew}) we have
\begin{eqnarray}
\dot{\rho}_D=-H\rho_D\left(2\frac{\sqrt{\Omega_D}}{na}+\frac{G'}{G}\right)\label{rhodot}.
\end{eqnarray}
where the prime stands for the derivative with respect to
$x=\ln{a}$. Inserting this equation in the conservation law
(\ref{consq}), we obtain the equation of state parameter
\begin{eqnarray}
w_D=-1+\frac{2}{3na}\sqrt{\Omega_D}+\frac{G'}{3G}\label{wDn}.
\end{eqnarray}
The equation of motion for $\Omega_D$ can be obtained by taking
the derivative of Eq. (\ref{Omegaqnew}). The result is
\begin{eqnarray}\label{Omegaq2new}
{\Omega'_D}=-\Omega_D\left(2\frac{\dot{H}}{H^2}+2\frac{\sqrt{\Omega_D}}{na
}\right).
\end{eqnarray}
where the dot is the derivative with respect to the time. The next
step is to calculate $\frac{\dot{H}}{H^2}$. Taking the derivative
of both side of the Friedman equation (\ref{Friednew}) with
respect to the cosmic time $t$, and using Eqs.
(\ref{Omega})-(\ref{consm}) and (\ref{wDn}), we obtain
\begin{eqnarray}\label{Hdotnew}
2\frac{\dot{H}}{H^2}=3(\Omega_D-1)-\frac{2}{na}\Omega^{3/2}_D-\Omega_k+\frac{G
'}{G}(1+\Omega_k-\Omega_D)
\end{eqnarray}
Substituting this relation into Eq. (\ref{Omegaq2new}), we obtain
the evolution behavior of the new agegraphic dark energy
\begin{eqnarray}\label{Omegaq3new}
{\Omega'_D}&=&\Omega_D\left[(1-\Omega_D)\left(3-\frac{2}{na}\sqrt{\Omega_D}\right)
+\Omega_k-\frac{G '}{G}(1+\Omega_k-\Omega_D)\right].
\end{eqnarray}
For completeness, we give the deceleration parameter
\begin{eqnarray}
q=-\frac{\ddot{a}}{aH^2}=-1-\frac{\dot{H}}{H^2},\label{qnew}
\end{eqnarray}
which combined with the Hubble parameter and the dimensionless
density parameters form a set of useful parameters for the
description of the astrophysical observations. Substituting Eq.
(\ref{Hdotnew}) in Eq. (\ref{qnew}) we get
\begin{eqnarray}\label{q}
q&=&\frac{1}{2}-\frac{3}{2}{\Omega_D} +\frac{\Omega^{3/2}_D}{na}
+\frac{\Omega_k}{2}-\frac{G '}{2G}(1+\Omega_k-\Omega_D).
\end{eqnarray}
\section{Interacting new ADE with varying gravitational constant\label{INT}} Next we consider the case
where the pressureless dark matter and the new ADE do not conserve
separately but interact with each other. Given the unknown nature of
both dark matter and dark energy there is nothing in principle
against their mutual interaction and it seems very special that
these two major components in the universe are entirely independent.
Indeed, this possibility is receiving growing attention in the
literature \cite{Ame,Zim, set} and appears to be compatible with
SNIa and CMB data \cite{Oli}. The total energy density satisfies a
conservation law
\begin{equation}\label{cons}
\dot{\rho}+3H(\rho+p)=0,
\end{equation}
where $\rho=\rho_{m}+\rho_{D}$ and $p=p_D$. However, as stated
above, both components- the pressureless dark matter and the new
ADE- are assumed to interact with each other; thus, one may grow
at the expense of the other. The conservation equations for them
read
\begin{eqnarray}
&& \dot{\rho}_D+3H\rho_D(1+w_D)=-Q,\label{consq2}\\
&&\dot{\rho}_m+3H\rho_m=Q, \label{consm2}
\end{eqnarray}
where $Q$ stands for the interaction term. Following \cite{Pav1}
we shall assume  for the latter the ansatz $Q =\Gamma\rho_D$ with
$\Gamma>0$ which means that there is an energy transfer from the
dark energy to dark matter. This expression for the interaction
term was first introduced in the study of the suitable coupling
between a quintessence scalar field and a pressureless cold dark
matter field \cite{Ame,Zim}. We also assume $\Gamma=3b^2(1+r)H$
where $r={\rho_m}/{\rho_D}$ and $b^2$ is a coupling constant.
Therefore, the interaction term $Q$ can be expressed as
\begin{eqnarray}\label{Q}
Q=3b^2H\rho_D(1+r),
\end{eqnarray}
where
\begin{eqnarray}\label{r}
r&=&\frac{\Omega_m}{\Omega_D} =-1+\frac{1+\Omega_k}{\Omega_D}.
\end{eqnarray}
Combining Eqs. (\ref{rhodot}), (\ref{Q}) and (\ref{r}) with  Eq.
(\ref{consq2}) we obtain the equation of state parameter
\begin{eqnarray}
w_D=-1+\frac{2}{3na}\sqrt{\Omega_D}-b^2\frac{(1+\Omega_k)}{\Omega_D}+\frac{G'}{3G}\label{wDnInt}.
\end{eqnarray}
If we take following \cite{jamil} $0<{G'}/{G} \leq 0.07$ and
assuming $\Omega_D= 0.73$ and $\Omega_k\approx 0.01$ for the
present time and $n=4$, $b=0.1$, we obtain $-0.87<w_D -\leq0.85$
which is consistent with recent observations. We can also find the
equation of motion for $\Omega_D$ by taking the derivative of Eq.
(\ref{Omegaqnew}). The result is
\begin{eqnarray}\label{Omegaq2new}
{\Omega'_D}=\Omega_D\left(-2\frac{\dot{H}}{H^2}-\frac{2}{na
}\sqrt{\Omega_D}\right).
\end{eqnarray}
where
\begin{eqnarray}\label{HdotInt}
2\frac{\dot{H}}{H^2}=3(\Omega_D-1)-\frac{2}{na}\Omega^{3/2}_D-\Omega_k-b^2\frac{(1+\Omega_k)}{\Omega_D}+\frac{G
'}{G}(1+\Omega_k-\Omega_D)
\end{eqnarray}
Substituting this relation into Eq. (\ref{Omegaq2new}), we obtain
the evolution behavior of the interacting new agegraphic dark
energy with variable gravitational constant
\begin{eqnarray}\label{OmegaqInt}
{\Omega'_D}&=&\Omega_D\left[(1-\Omega_D)\left(3-\frac{2}{na}\sqrt{\Omega_D}\right)
+\Omega_k-3b^2(1+\Omega_k)-\frac{G
'}{G}(1+\Omega_k-\Omega_D)\right].
\end{eqnarray}
The deceleration parameter is now given by
\begin{eqnarray}\label{qInt}
q&=&\frac{1}{2}-\frac{3}{2}{\Omega_D} +\frac{\Omega^{3/2}_D}{na}
+\frac{\Omega_k}{2}-\frac{3}{2}b^2(1+\Omega_k)-\frac{G
'}{2G}(1+\Omega_k-\Omega_D).
\end{eqnarray}
Again takeing  $0<{G'}/{G} \leq 0.07$ and assuming $\Omega_D=
0.73$ and $\Omega_k\approx 0.01$ for the present time and $n=4$,
$b=0.1$, we obtain $-0.46<q -\leq0.45$ which is again compatible
with recent observational data \cite{Daly}.
\section{Interacting viscous new ADE with varying $G$\label{INT}}

In this section we would like to generalize our study to the
interacting viscous new agegraphic dark energy model. In an
isotropic and homogeneous FRW universe, the dissipative effects
arise due to the presence of bulk viscosity in cosmic fluids. The
theory of bulk viscosity was initially investigated by Eckart
\cite{Eck1} and later on pursued by Landau and Lifshitz
\cite{Lan}. Dark energy with bulk viscosity has a peculiar
property to cause accelerated expansion of phantom type in the
late evolution of the universe \cite{Bre1,Bre2,Bre3}. It can also
alleviate several cosmological puzzles like age problem,
coincidence problem  and phantom crossing. The energy-momentum
tensor of the viscous fluid is
\begin{equation}\label{T}
T_{\mu\nu}=\rho_Du_{\mu}u_{\nu}+\tilde{p}_D(g_{\mu\nu}+u_{\mu}u_{\nu}),
\end{equation}
where $u_{\mu}$ is the four-velocity vector and
\begin{equation}\label{consv}
\tilde{p}_D={p}_D-3H\xi,
\end{equation}
is the effective pressure of dark energy and $\xi$ is the bulk
viscosity coefficient. We require $\xi>0$ to get positive entropy
production in conformity with second law of thermodynamics
\cite{Z}. The energy conservation equation for interacting viscous
dark energy is now given by
\begin{eqnarray}
\dot{\rho}_D+3H(\rho_D+\tilde{p}_D)=-Q,\label{consqv}
\end{eqnarray}
which can be written
\begin{eqnarray}
\dot{\rho}_D+3H\rho_D(1+w_D)=9H^2\xi-Q,\label{consqv2}
\end{eqnarray}
Combining Eqs. (\ref{rhodot}), (\ref{Q}) and (\ref{r}) with  Eq.
(\ref{consqv2}) we obtain the equation of state parameter
\begin{eqnarray}
w_D=-1+\frac{2}{3na}\sqrt{\Omega_D}-b^2\frac{(1+\Omega_k)}{\Omega_D}+\frac{G'}{3G}+\frac{3H\xi}{\rho_D}\label{wDnIntv}.
\end{eqnarray}
If we assume $\xi=\alpha H^{-1}\rho_D$, where $\alpha$ is a constant
parameter, then we get
\begin{eqnarray}
w_D=3\alpha-1+\frac{2}{3na}\sqrt{\Omega_D}-b^2\frac{(1+\Omega_k)}{\Omega_D}+\frac{G'}{3G}\label{wDnIntv2}.
\end{eqnarray}
The equation of motion for viscous ADE is obtained as
\begin{eqnarray}\label{OmegaqIntv}
{\Omega'_D}&=&\Omega_D\left[(1-\Omega_D)\left(3-\frac{2}{na}\sqrt{\Omega_D}\right)
+\Omega_k-3b^2(1+\Omega_k)-\frac{G
'}{G}(1+\Omega_k-\Omega_D)+9\alpha\Omega_D\right].
\end{eqnarray}

\section{Conclusions\label{CONC}}
In this work we have investigated the interacting new agegraphic
viscous dark energy scenario with a varying gravitational constant.
 We have obtained the
equation of state and the deceleration parameters for both
interacting and noninteracting new agegraphic dark energy. By
consideing non-interacting and interacting cases we have extracted
the exact differential equations that determine the evolution of
the dark energy density-parameter, where the $G$-variation
appears as a coefficient in additional terms.

\section{Acknowledgment}This work has been supported by Research Institute for Astronomy
and Astrophysics of Maragha, Iran.


\begin{thebibliography}{99}
\bibitem{1}
A. G.  Riess {\it{et al.}} [Supernova Search Team Collaboration],
Astrophys. J. {\bf 607}, 665 (2004) [arXiv:astro-ph/0402512];
 R. A. Knop
{\it{et al.}}, [Supernova Cosmology Project Collaboration],
Astrophys. J. {\bf 598}, 102 (2003) [arXiv:astro-ph/0309368]; S.
Perlmutter {\it{et al.}} [Supernova Cosmology Project
Collaboration], Astrophys. J. {\bf 517}, 565 (1999)
[arXiv:astro-ph/9812133].

\bibitem{2}
C. L. Bennett {\it{et al.}}, Astrophys. J. Suppl. {\bf 148}, 1
(2003) [arXiv:astro-ph/0302207]; D. N. Spergel {\it{et al.}},
Astrophys. J. Suppl. {\bf 148}, 175 (2003) [arXiv:astro-ph/0302209].

\bibitem{3}
M. Tegmark {\it{et al.}} [SDSS Collaboration], Phys. Rev. D {\bf
69}, 103501 (2004) [arXiv:astro-ph/0310723]; U. Seljak {\it{et
al.}}, Phys. Rev. D {\bf 71}, 103515 (2005) [astro-ph/0407372]; J.
K. Adelman-McCarthy {\it{et al.}} [SDSS Collaboration],
[arXiv:astro-ph/0507711].

\bibitem{4}
S. W. Allen, {\it{et al.}}, Mon. Not. Roy. Astron. Soc. {\bf 353},
457 (2004) [astro-ph/0405340].

\bibitem{c7}
V. Sahni and A. Starobinsky, Int. J. Mod. Phy. D {\bf 9}, 373
(2000); P. J. Peebles and B. Ratra, Rev. Mod. Phys. {\bf 75}, 559
(2003).

\bibitem{8}P.~J.~Steinhardt,  {\it {Critical Problems in Physics}} (1997), Princeton
University Press.

\bibitem{quint}
B.~Ratra and P.~J.~E.~Peebles, Phys.\ Rev.\ D {\bf 37}, 3406 (1988);
C.~Wetterich, Nucl.\ Phys.\ B {\bf 302}, 668 (1988); A.~R.~Liddle
and R.~J.~Scherrer, Phys.\ Rev.\ D {\bf 59}, 023509 (1999);
I.~Zlatev, L.~M.~Wang and P.~J.~Steinhardt, Phys.\ Rev.\ Lett.\ {\bf
82}, 896 (1999).
\bibitem{phant} R. R. Caldwell, Phys.
Lett. B {\bf{545}}, 23 (2002); R.~R.~Caldwell, M.~Kamionkowski and
N.~N.~Weinberg, Phys. Rev. Lett. {\bf 91}, 071301 (2003); S. Nojiri
and S. D. Odintsov, Phys. Lett. B {\bf 562}, 147 (2003); V. K.
Onemli and R. P. Woodard, Phys.\ Rev.\ D {\bf 70}, 107301 (2004); M.
R. Setare, J. Sadeghi, A. R. Amani, Phys. Lett. B {\bf 666}, 288,
(2008);
  M.~R.~Setare and E.~N.~Saridakis,
  JCAP {\bf 0903}, 002 (2009).
 %

\bibitem{quintom}
B.~Feng, X.~L.~Wang and X.~M.~Zhang, Phys.\ Lett.\  B {\bf 607}, 35
(2005);
Z. K. Guo, {\it{et al.}}, Phys. Lett. B {\bf 608}, 177 (2005); M.-Z
Li, B. Feng, X.-M Zhang, JCAP, 0512, 002 (2005);  M. R. Setare,
Phys. Lett. B {\bf 641}, 130 (2006);  M. R. Setare, J. Sadeghi, and
A. R. Amani, Phys. Lett. B {\bf 660}, 299 (2008);
  M.~R.~Setare and E.~N.~Saridakis,
  Phys.\ Lett.\  B {\bf 668}, 177 (2008);
  M.~R.~Setare and E.~N.~Saridakis,
  JCAP {\bf 0809}, 026 (2008);
  M.~R.~Setare and E.~N.~Saridakis,
  Int.\ J.\ Mod.\ Phys.\  D {\bf 18}, 549 (2009).
  \bibitem{hologram} L. Susskind, J. Math. Phys, {\bf36}, (1995),
6377-6396.
\bibitem{cohen} A. Cohen, D. Kaplan and A. Nelson, Phys. Rev. Lett
82, (1999), 4971.
\bibitem{cai1}R. G. Cai, Phys. Lett. B {\bf 657}, 228, (2007).
\bibitem{zin}I. P. Neupane, Phys. Lett. B {\bf 673}, 111, (2009).
\bibitem{kar}F. Karolyhazy, Nuovo.Cim. A 42, 390 (1966);\\
F. Karolyhazy, A. Frenkel and B. Lukacs, in {\it Physics as natural
Philosophy}\\  edited by A. Shimony and H. Feschbach, MIT Press,
Cambridge, MA, (1982);\\ F. Karolyhazy, A. Frenkel and B. Lukacs, in
{\it Quantum Concepts in Space and Time}\\  edited by R. Penrose and
C.J. Isham, Clarendon Press, Oxford, (1986).

\bibitem{maz}M. Maziashvili, Int. J. Mod. Phys. D {\bf 16} (2007) 1531;
M. Maziashvili, Phys. Lett. B {\bf652} (2007) 165.



\bibitem{age} H. Wei and R. G. Cai, Eur. Phys. J. C 59 (2009) 99;\\ H.
Wei and R. G. Cai, Phys. Lett. B 663 (2008) 1;
\\ J. Cui, et al., arXiv:0902.0716; \\ Y. W. Kim, et al.,  Mod. Phys. Lett. A 23 (2008) 3049;\\
Y. Zhang, et al. arXiv:0708.1214;\\ J .P Wu, D. Z. Ma, Y. Ling,
Phys.
Lett. B 663,  (2008) 152; \\ K. Y. Kim, H. W. Lee, Y. S. Myung, Phys.Lett. B 660 (2008) 118;\\
X. Wu, et al., arXiv:0708.0349;\\ J. Zhang, X. Zhang, H. Liu, Eur.
Phys. J. C 54 (2008) 303; \\ I. P. Neupane, Phys. Lett. B 673
(2009) 111.
\bibitem{shey1} A. Sheykhi, Phys. Lett. B 680 (2009) 113.
\bibitem{shey2} A. Sheykhi, Phys. Lett. B 682 (2010) 329; 
\\ A. Sheykhi, Int. J. Mod. Phys. D 18, No. 13 (2009) 2023; 
\\ A. Sheykhi, Phys. Rev. D 81 (2010) 023525; 
\\ A. Sheykhi, Int. J. Mod. Phys. D Vol. 19, No. 3 (2010) 305. 
 \\ A. Sheykhi, arXiv:0909.0302. 


\bibitem{setare} M. R. Setare, arXiv:0907.4910;\\ M. R. Setare,
arXiv:0908.0196.

\bibitem{age2} H. Wei and R. G. Cai, Phys. Lett. B 663 (2008) 1.

\bibitem{Bertolami:2007zm}
  O.~Bertolami, F.~Gil Pedro and M.~Le Delliou,
  Phys.\ Lett.\  B {\bf 654}, 165 (2007)
  [arXiv:astro-ph/0703462];\\
  M.~Le Delliou, O.~Bertolami and F.~Gil Pedro,
  AIP Conf.\ Proc.\  {\bf 957}, 421 (2007)
  [arXiv:0709.2505 [astro-ph]].
\bibitem{Feng:2007wn}
  C.~Feng, B.~Wang, Y.~Gong and R.~K.~Su,
  JCAP {\bf 0709}, 005 (2007)
  [arXiv:0706.4033 [astro-ph]];\\
   E.~Abdalla, L.~R.~W.~Abramo, L.~.~J.~Sodre and B.~Wang,
  ``Signature of the interaction between dark energy and dark matter in galaxy clusters,''
  arXiv:0710.1198 [astro-ph].
  \bibitem{shee}P. Coles
 and F. Lucchin , Cosmology: The origin and evolution of cosmic
structure (John Wiley, 2003).
\bibitem{Bre1} I. Brevik and O. Gorbunova, Gen. Relativ. Gravit. 37, 2039
(2005).
\bibitem{lanc}P. Langacher,   Phys. Rep. 72 185, (1981).


\bibitem{jamil}
M. Jamil, E. N. Saridakis, M. R. Setare, Phys. Lett. B 679, 172,
(2009).

\bibitem{manos} J. Lu, E. N. Saridakis,  M. R. Setare,  L. Xu,  JCAP 1003, 031,
(2010).
  \bibitem{set1}M. R. Setare, Phys. Lett. B 642 (2006) 1.
\bibitem{Wei2} H. Wei and R. G. Cai, Phys. Lett. B 660 (2008) 113.

\bibitem{spe} C. L. Bennett, et al.,  Astrophys. J. Suppl.
148 (2003) 1;\\ D. N. Spergel, Astrophys. J. Suppl. 148 (2003) 175;\\
M. Tegmark, et al., Phys. Rev. D 69 (2004) 103501;\\ U. Seljak, A.
Slosar, P. McDonald, JCAP 0610 (2006) 014;\\ D. N. Spergel, et
al., Astrophys. J. Suppl. 170 (2007) 377.


\bibitem{Ame} L. Amendola, Phys. Rev. D 60 (1999)  043501; \\ L. Amendola, Phys. Rev. D 62 (2000) 043511;
 \\ L. Amendola and C. Quercellini, Phys. Rev. D 68
(2003)  023514; \\ L. Amendola and D. Tocchini-Valentini, Phys. Rev.
D 64 (2001)  043509 .

\bibitem{Zim} W. Zimdahl and D. Pavon, Phys. Lett. B 521 (2001) 133;\\ W. Zimdahl and D. Pavon, Gen. Rel. Grav. 35
(2003) 413;\\ L. P. Chimento, A. S. Jakubi, D. Pavon and W.
Zimdahl, Phys. Rev. D 67 (2003)  083513.
\bibitem{set}M. R. Setare, Eur. Phys. J. C 50 (2007) 991;\\
M. R. Setare, JCAP 0701 (2007) 023;\\ M. R. Setare, Phys. Lett. B
654 (2007) 1;\\ M. R. Setare, Phys. Lett. B 642 (2006) 421.
\bibitem{Oli} G. Olivares, F. Atrio, D. Pavon, Phys. Rev. D 71 (2005) 063523.

\bibitem{Pav1} D. Pavon, W. Zimdahl, Phys. Lett. B 628 (2005) 206.


\bibitem{Eck1} C. Eckart, Phys. Rev. 58 (1940) 919.

\bibitem{Lan} L.D. Landau and E.M. Lifshitz, Fluid Mechanics (Butterworth
Heineman, 1987)



\bibitem{Bre2} I. Brevik, O. Gorbunova and Y. A. Shaido, Int. J. Mod.
Phys. D 14, 1899 (2005);\\
I. Brevik and O. Gorbunova, Eur. Phys. J. C 56, 425 (2008).

\bibitem{Bre3} I. Brevik, O. Gorbunova, D. S. Gomez,
arXiv:0908.2882.

\bibitem{Z} W. Zimdahl and D. Pavon, Phys. Rev. D 61 (2000) 108301;
\bibitem{kogan}
  G.~S.~Bisnovatyi-Kogan,
  Int.\ J.\ Mod.\ Phys.\  D {\bf 15}, 1047 (2006).


\bibitem{Damour}
Damour T.,{\it  et al}, Phys. Rev. Lett. {\bf 61}, 1151 (1988).

\bibitem{guenther} D.B. Guenther, Phys. Lett. B {\bf 498}, 871 (1998).

\bibitem{Gaztanaga}
  E.~Gaztanaga, E.~Garcia-Berro, J.~Isern, E.~Bravo and I.~Dominguez,
  Phys.\ Rev.\  D {\bf 65}, 023506 (2002).


\bibitem{Biesiada} Biesiada M. and Malec B.,  Mon. Not. R. Astron. Soc. {\bf 350}, 644
(2004).



\bibitem{ray1}
  S.~Ray and U.~Mukhopadhyay,
  Int.\ J.\ Mod.\ Phys.\  D {\bf 16}, 1791 (2007).

\bibitem{Ber} P.G. Bergmann, Int. J. Theor. Phys. 1 (1968) 25; R.V.
Wagoner, Phys. Rev. D 1 (1970) 3209; K. Nordtvedt, Astrophys. J.
161 (1970) 1059.

\bibitem{Bek} J.D. Bekenstein, Found. Phys. 16 (1986) 409.

\bibitem{Shap} I.L. Shapiro, J. Sola, JHEP 0202 (2002) 006; A. Babi ´c, B.
Guberina, R. Horvat, H. Štefan¡ci ´ c, Phys. Rev. D 65 (2002)
085002; I.L. Shapiro, J. Sola, C. Espana-Bonet, P. Ruiz-Lapuente,
Phys. Lett. B 574 (2003) 149.


\bibitem{Reu} M. Reuter, Phys. Rev. D 57 (1998) 971; A. Bonnano, M. Reuter,
Phys. Rev. D 65 (2002) 043508.

\bibitem{Horv} R. Horvat, Phys. Rev. D 70 (2004) 087301.

\bibitem{Gub} B. Guberina, R. Horvat, H. Nikolic, Phys. Rev. D 72 (2005)
125011.
\bibitem{Daly} R.A. Daly et al., Astrophysics J. 677 (2008) 1.

\end{thebibliography}
\end{document}